\documentclass{article}
\usepackage{spconf,amsmath,graphicx}

\usepackage{amssymb}
\usepackage{bm}
\usepackage{siunitx}
\usepackage{todonotes}
\usepackage{booktabs}
\usepackage{subcaption}
\usepackage{adjustbox}

\newcommand{\by}{\text{x}}

\newcommand{\tabvspace}{
\midrule
}
\newcommand{\mathg}{G}
\newcommand{\mathd}{D}

\newcommand{\norm}[1]{\lVert#1\rVert}
\newcommand{\wer}[1]{$#1\%$}

\title{Exploring Speech Enhancement with Generative Adversarial Networks for Robust Speech Recognition}
\twoauthors
 {Chris Donahue\sthanks{Work performed as an intern at Google.}\sthanks{Copyright (c) 2017 IEEE. Personal use of this material is permitted. However, permission to reprint/republish this material for advertising or promotional purposes or for creating new collective works for resale or redistribution to servers or lists, or to reuse any copyrighted component of this work in other works must be obtained from the IEEE.}}
	{
    	UC San Diego Department of Music \\
      \fontsize{9}{9}\selectfont\ttfamily\upshape
      cdonahue@ucsd.edu
    }
 {Bo Li, Rohit Prabhavalkar}
	{
    	Google \\
      \fontsize{9}{9}\selectfont\ttfamily\upshape
      \{boboli,prabhavalkar\}@google.com
    }
\begin{document}
%
\maketitle
\begin{abstract}

We investigate the effectiveness of generative adversarial networks (GANs) for speech enhancement,
in the context of improving noise robustness of automatic speech recognition (ASR) systems.
Prior work~\cite{pascual2017segan} demonstrates that GANs can effectively suppress additive noise in raw waveform speech signals,
improving perceptual quality metrics;
however this technique was not justified in the context of ASR.
In this work,
we conduct a detailed study to measure the effectiveness of GANs in enhancing speech contaminated by both additive and reverberant noise.
Motivated by recent advances in image processing~\cite{isola2016image},
we propose operating GANs on log-Mel filterbank spectra instead of waveforms,
which requires less computation and is more robust to reverberant noise.
While GAN enhancement improves the performance of a clean-trained ASR system on noisy speech,
it falls short of the performance achieved by conventional multi-style training (MTR).
By appending the GAN-enhanced features to the noisy inputs and retraining,
we achieve a $7$\% WER improvement relative to the MTR system.

\end{abstract}

\section{Introduction}
\label{sec:intro}

Speech enhancement techniques aim to improve the quality of speech by reducing noise.
They are crucial components, either explicitly~\cite{chen2008fundamentals} or implicitly~\cite{xu2014experimental, li2016neural}, in ASR systems for noise robustness.
Even with state-of-the-art deep learning-based ASR models, noise reduction techniques can still be beneficial~\cite{li2017improved}.
Besides the conventional enhancement techniques~\cite{chen2008fundamentals}, deep neural networks have been widely adopted to either directly reconstruct clean speech~\cite{feng2014speech, xu2015regression} or estimate masks~\cite{wang2017supervised, li2014spectral, narayanan2013ideal} from the noisy signals.
Different types of networks have also been investigated in the literature for enhancement, such as denoising autoencoders~\cite{lu2013speech}, convolution networks~\cite{fu2016snr} and recurrent networks~\cite{weninger2015speech}.

In their limited history,
GANs~\cite{goodfellow2014generative} have attracted attention for their ability to synthesize convincing images when trained on corpora of natural images.
Refinements to network architecture have improved the fidelity of the synthetic images~\cite{radford2015unsupervised}.
Isola et al.~\cite{isola2016image} demonstrate the effectiveness of GANs for image ``translation'' tasks,
mapping images in one domain to related images in another.
In spite of the success of GANs for image synthesis,
exploration on audio has been limited.
Pascual et al.~\cite{pascual2017segan} demonstrate promising performance of GANs for speech enhancement in the presence of additive noise,
posing enhancement as a translation task from noisy signals to clean ones.
Their method,
speech enhancement GAN (SEGAN),
yields improvements to perceptual speech quality metrics over the noisy data and traditional enhancement baselines.
Their investigation seeks to improve speech quality for telephony rather than ASR.

In this work,
we study the benefit of GAN-based speech enhancement for ASR.
In order to limit the confounding factors in our study,
we use an existing ASR model trained on clean speech data to measure the effectiveness of GAN-based enhancement.
To gauge performance under more-realistic ASR conditions,
we consider reverberation in addition to additive noise.
We first train a SEGAN model to map simulated noisy speech to the original clean speech in the time domain.
Then, we measure the performance of the ASR model on noisy speech before and after enhancement by SEGAN.
Our experiment indicates that SEGAN does not improve ASR performance under these noise conditions.

\begin{figure*}
  \centering
  \includegraphics[width=0.85\linewidth]{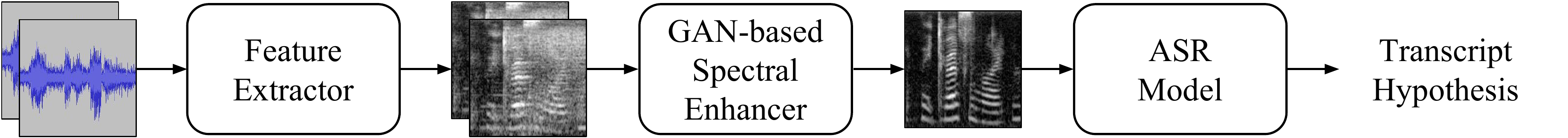}
  \caption{System overview.}
  \label{fig:system}
  \vspace{-0.1in}
\end{figure*}

To address this, 
we refine the SEGAN method to operate on a time-frequency representation, 
specifically, log-Mel filterbank spectra. 
With this \emph{spectral feature mapping} (SFM) approach, 
we can pass the output of our enhancement model directly to the ASR model (Figure~\ref{fig:system}). 
While deep learning has previously been applied to SFM for ASR~\cite{maas2012recurrent,han2015deep,narayanan2015improving}, 
our work is the first to use GANs for this task. 
Michelsanti et al.~\cite{michelsanti2017conditional} employ GANs for SFM, 
but target speaker verification rather than ASR. 
Our frequency-domain approach improves ASR performance dramatically,
though performance is comparable to the same enhancement model trained with an L1 reconstruction loss. 
Anecdotally speaking, 
the GAN-enhanced spectra appear more realistic than the L1-enhanced spectra when visualized (Figure \ref{fig:fsegan_vis}),
suggesting that 
ASR models may not benefit from the fine-grained details that GAN enhancement produces. 

State-of-the-art ASR systems use MTR~\cite{lippmann1987multi} to achieve robustness to noise at inference time.
While this strategy is known to be effective,
a resultant model may still benefit from enhancement as a preprocessing stage.
To measure this effect,
we also use an existing ASR model trained with MTR
and compare its performance on noisy speech with and without enhancement.
We find that GAN-based enhancement degrades performance of this model,
even with retraining.
However, retraining the MTR model with both noisy and enhanced features in its input representation
improves performance.

\section{Generative Adversarial Networks}

Generative adversarial networks (GANs) are unsupervised generative models that
learn to produce realistic samples of a given dataset from low-dimensional, random latent vectors~\cite{goodfellow2014generative}.
GANs consist of two models (usually neural networks),
a \emph{generator} and a \emph{discriminator}.
The generator $\mathg$ maps latent vectors drawn from some known prior $p_{z}$ to samples:
$\mathg~:~\bm{z}~\mapsto~\bm{\hat{y}}$, where $\bm{z}~\sim~p_{z}$.
The discriminator $\mathd$ is tasked with determining if a given sample is real
($\bm{y}~\sim~p_{data}$, a sample from the real dataset)
or fake
($\mathg(\bm{z})~\sim~p_{G}$, where $p_{G}$ is the implicit distribution of the generator when
$\bm{z}~\sim~p_{z}$).
The two models are pitted against each other in an adversarial framework.


Real-world datasets often contain additional information associated with each example,
e.g. the type of object depicted in an image.
Conditional GANs (cGANs)~\cite{mirza2014conditional} use this information $\bm{x}$ by providing it as input to the generator, typically in a one-hot representation:
$\mathg : \{\bm{x}, \bm{z}\} \mapsto \bm{\hat{y}}$.
After training,
we can sample from the generator's implicit posterior $p_{G}(\bm{\hat{y}} \mid \bm{x})$ by fixing $\bm{x}$ and sampling $\bm{z} \sim p_{z}$.
To accomplish this,
$\mathg$ is trained to minimize the following objective, while $\mathd$ is trained to maximize it:
\begin{multline}
\vspace{-0.2in}
\label{eq:cgan}
\mathcal{L}_{cGAN}(\mathg, \mathd) =
  \mathbb{E}_{\bm{x,y} \sim p_{data}}[\log \mathd(\bm{x}, \bm{y})] + \\
  \mathbb{E}_{\bm{x} \sim p_{data}, \bm{z} \sim p_{z}}[\log (1 - \mathd(\bm{x}, \mathg(\bm{x}, \bm{z})))].
\end{multline}

Recently, researchers have used full-resolution images as conditioning information. Isola et al.~\cite{isola2016image} propose a cGAN approach to address image-to-image ``translation'' tasks,
where appropriate datasets consist of matched pairs of images $(\bm{x}, \bm{y})$ in two different domains.
Their approach,
dubbed \emph{pix2pix},
uses a convolutional generator that receives as input an image $\bm{x}$, a latent vector $\bm{z}$ and produces $\mathg(\bm{x}, \bm{z})$: an image of identical resolution to $\bm{x}$.
A convolutional discriminator is shown pairs of images stacked along the channel axis and is trained to determine if the pair is real $(\bm{x}, \bm{y})$ or fake $(\bm{x}, \mathg(\bm{x},\bm{z}))$.

For conditional image synthesis,
prior work~\cite{pathak2016context} demonstrates the effectiveness of combining the GAN objective with an unstructured loss. 
Noting this, Isola et al.~\cite{isola2016image} use a hybrid objective to optimize their generator,
penalizing it for L1 reconstruction error in addition to the adversarial objective:
\begin{multline}
\vspace{-0.2in}
\label{eq:pix2pix}
\min_{\mathg} \max_{\mathd} V(\mathg, \mathd) =
\mathcal{L}_{cGAN}(\mathg, \mathd) + 100 \cdot \mathcal{L}_{L1}(\mathg), \\
\text{where }
\mathcal{L}_{L1}(\mathg) = \mathbb{E}_{\bm{x,y} \sim p_{data}, \bm{z} \sim p_{z}}[\norm{\bm{y} - \mathg(\bm{x},\bm{z})}_1].
\end{multline}



\section{Method}

We describe our approach to spectral feature mapping using GANs,
beginning by outlining the related time-domain SEGAN approach.

\subsection{SEGAN}
\label{sec:segan}

Pascual et al.~\cite{pascual2017segan} propose SEGAN,
a technique for enhancing speech in the time domain.
The SEGAN method is a 1D adaptation of the 2D pix2pix~\cite{isola2016image} approach.
The fully-convolutional generator receives second-long ($16384$ samples at $\SI{16}{\kilo\hertz}$) windows of noisy speech as input and is trained to output clean speech.
During inference,
the generator is used to enhance longer segments of speech by repeated application on one-second windows without overlap.

The generator's encoder consists of $11$ layers of stride-$2$ convolution with increasing depth,
resulting in a feature map at the bottle-neck of $8$ timesteps with depth $1024$.
Here, the authors append a latent noise vector $\bm{z}$ of the same dimensionality along the channel axis.
The resultant $8 \by 2048$ matrix is input to an $11$-layer upsampling decoder,
with skip connections from corresponding input feature maps.
As a departure from pix2pix,
the authors remove batch normalization from the generator.
Furthermore, they use 1D filters of width $31$ instead of 2D filters of size $4 \by 4$.
They also substitute the traditional GAN loss function with the least squares GAN objective~\cite{mao2016least}.

In agreement with observations from~\cite{mathieu2015deep},
we found that the SEGAN generator learned to ignore $\bm{z}$.
We hypothesize that latent vectors may be unnecessary given the presence of noise in the input.
We removed the latent vector from the generator altogether;
the resultant deterministic model demonstrated improved performance in our experiments.




\subsection{FSEGAN}
\label{sec:fsegan}

It is common practice in ASR to preprocess time-domain speech data into time-frequency \emph{spectral} representations.
Phase information of the signal is typically discarded; 
hence, an enhancement model only needs to reconstruct the magnitude information of the time-frequency representation.
We hypothesize that performing GAN enhancement in the frequency domain
will be more effective than in the time domain. 
With this motivation, we propose a frequency-domain SEGAN (FSEGAN),
which performs spectral feature mapping using an approach similar to pix2pix~\cite{isola2016image}. 
FSEGAN ingests time-windowed spectra of noisy speech and is trained to output clean speech spectra.

\begin{figure}
  \centering
  \includegraphics[width=0.6\linewidth]{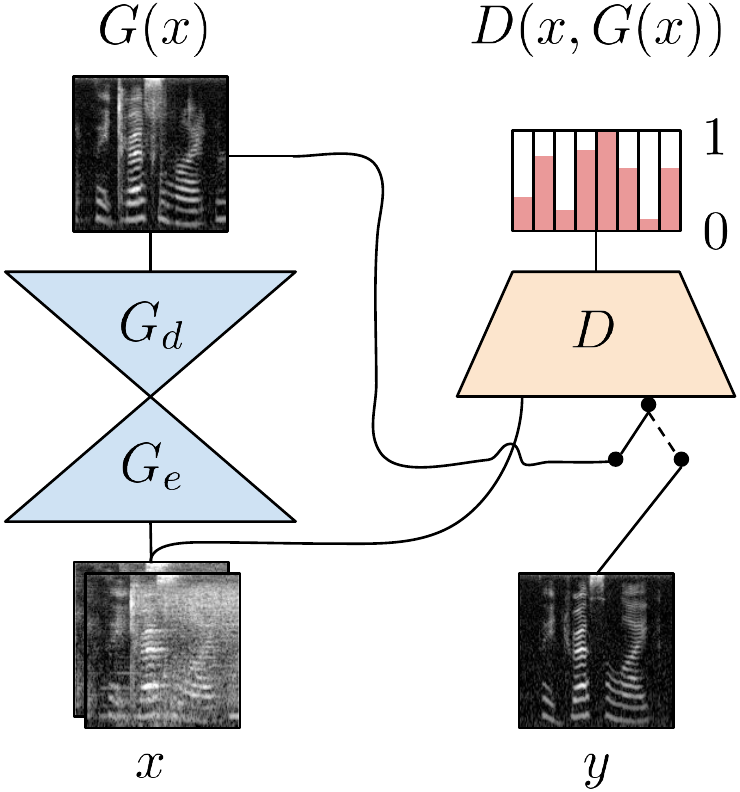}
  \caption{
  Time-frequency FSEGAN enhancement strategy.
  $\mathg$~(composition of encoder $\mathg_{e}$ and decoder $\mathg_{d}$)
  maps stereo, noisy spectra $\bm{x}$ to enhanced $\mathg(\bm{x})$.
  $\mathd$ receives as input either $(\bm{x}, \bm{y})$ or $(\bm{x}, \mathg(\bm{x}))$
  and decides if the pair is real or enhanced.
  }
  \label{fig:fsegan}
  \vspace{-0.2in}
\end{figure}

The fully-convolutional FSEGAN generator contains $7$ encoder and $7$ decoder layers 
($4 \by 4$ filters with stride $2$ and increasing depth), 
and features skip connections across the bottleneck between corresponding layers. 
The final decoder layer has linear activation and outputs a single channel. 
As with SEGAN, 
we exclude both batch normalization and latent codes $\bm{z}$ from our generator, 
resulting in a deterministic model. 
The discriminator contains $4$ convolutional layers with $4 \by 4$ filters and a stride of $2$.
A final $1 \by 8$ layer (stride $1$ with \emph{sigmoid} nonlinearity) 
aggregates the activations from $8$ frequency bands into a single decision for each of $8$ timesteps. 
We train FSEGAN with the objective in Equation~\ref{eq:pix2pix}. 
Other architectural details are identical to pix2pix~\cite{isola2016image}.\footnote{We modify the following open-source implementation: \texttt{https://github.com/affinelayer/pix2pix-tensorflow}}
The FSEGAN approach is depicted in Figure~\ref{fig:fsegan}.

\section{Experiments}


\subsection{Dataset}
\label{sec:data}


We use the Wall Street Journal (WSJ) corpus~\cite{paul1992design} as our source of clean speech data.
Specifically, we train on the $\SI{16}{\kilo\hertz}$, speaker-independent SI-284 set ($81$ hours, $284$ speakers, 37k utterances).
We perform validation on the dev93 set and evaluate on the eval92 set.

We use large, stereo datasets of musical and ambient signals as our additive noise sources for MTR.
This data is collected from YouTube and recordings of daily life environments.
During training,
we use discrete mixtures ranging from $\SI{0}{\decibel}$ to $\SI{30}{\decibel}$ SNR, averaging $\SI{11}{\decibel}$.
At test time,
the SNRs are slightly offset,
ranging from $\SI{0.2}{\decibel}$ to $\SI{30.2}{\decibel}$.

As a source of reverberation for MTR,
we use a room simulator as described in~\cite{kim2017generation}.
The simulator randomizes the positions of the speech and noise sources,
the position of a virtual stereo microphone,
the T$60$ of the reverberation,
and the room geometry.
Through this process, our monaural speech data becomes stereo.
Room configurations for training and testing are drawn from distinct sets;
they are randomized during training and fixed during testing.

\subsection{ASR Model}
\label{sec:asr}

We train a monaural listen, attend and spell (LAS) model~\cite{chan2015listen} on the clean WSJ training data as described in Section~\ref{sec:data},
performing early stopping by the WER of the model on the validation set.
To compare the effectiveness of GAN-based enhancement to MTR,
we also train the same model using MTR as described in Section~\ref{sec:data},
using only one channel of the noisy speech.
We refer to the clean-trained model as ASR-Clean and the MTR-trained model as ASR-MTR.

To preprocess the time-domain data,
we first apply the short-time Fourier transform with a window size of $\SI{32}{\milli\second}$ and a hop size of $\SI{10}{\milli\second}$.
We retain the magnitude spectrum of the output and discard the phase.
Then, we calculate triangular windows for a bank of $128$ filters, 
where filter center frequencies are equally spaced on the Mel scale between $\SI{125}{\hertz}$ and $\SI{7500}{\hertz}$.
After applying this transform to the magnitude spectrum,
we take the logarithm of the output and normalize each frequency bin to have zero mean and unit variance.

To process these features,
our LAS encoder contains two convolutional layers with filter sizes: 1)~$3\by{}5\by{}1\by{}32$, 
and 2)~$3\by{}3\by{32}\by{32}$.
The activations of the second layer are passed to a bidirectional, convolutional LSTM layer~\cite{zhang2017very,xingjian2015convolutional},
followed by three bidirectional LSTM layers.
The decoder contains a unidirectional LSTM with additive attention~\cite{bahdanau2014neural}
whose outputs are fed to a \emph{softmax} over characters.


\subsection{GAN}

For our GAN experiments,
we generate multi-style, matched pairs of noisy and clean speech in the manner described in Section~\ref{sec:data}.
For our FSEGAN experiments,
we transform these pairs into time-frequency spectra in a manner identical to that of the ASR model described in Section~\ref{sec:asr}.
We frame the pairs into $\SI{1.28}{\second}$ windows with $50\%$ overlap
and train with random minibatches of size $100$.
The resultant SEGAN inputs are $20480$ samples long and
the FSEGAN inputs are $128\by{}128$.
In alignment with~\cite{pascual2017segan},
we use no overlap during evaluation.
We perform early stopping based on the WER of ASR-Clean on the enhanced validation set.

\section{Results}

\begin{table}[!t]
\small
\centering

\begin{tabular}{llrr}

    \toprule
    Test Set   		  & Enhancer & ASR-Clean WER        & ASR-MTR WER          \\
    \midrule
    Clean             & None    & $11.9$      & $14.3$        \\
    \tabvspace
    MTR  & None    & $72.2$      & $20.3$        \\
      	   & SEGAN   & $80.7$      & $52.8$        \\
           & FSEGAN  & $33.3$      & $25.4$        \\
    \bottomrule

\end{tabular}
\caption{Results of GAN enhancement experiments.}
\label{tab:results}
\vspace{-0.2in}
\end{table}

We compute the WER of ASR-Clean and ASR-MTR on both the clean and MTR test sets.
We also compute the WER of both models on the MTR test set enhanced by SEGAN and FSEGAN.
Results are shown in Table \ref{tab:results}.
While the WER of ASR-Clean (\wer{11.9}) is not state-of-the-art,
we focus more on relative performance with enhancement.
Our previous work~\cite{prabhavalkar2017comparison} has shown that LAS can approach state-of-the-art performance when trained on larger amounts of data.

The SEGAN method degrades performance of ASR-Clean on the MTR test set by
\wer{12} relative. 
To verify the accuracy of this result,
we also ran an experiment to remove only additive noise with SEGAN:
the conditions in the original paper.
Under that condition, 
we found that 
SEGAN improved performance of ASR-Clean by
\wer{21} 
relative,
indicating that SEGAN struggles to suppress reverberation.

In contrast,
our FSEGAN method improves the performance of ASR-Clean by \wer{54} relative.
While this is a dramatic improvement,
it does not exceed the performance achieved with MTR training (\wer{33} vs. \wer{20} WER).
Furthermore,
FSEGAN degraded performance for ASR-MTR, consistent with observations in~\cite{narayanan2014joint}.


\begin{figure}[!t]
  \centering
\vspace{2.5mm}
  \begin{subfigure}[b]{0.47\linewidth}
 	\centering
    \includegraphics[trim={220 5 910 20},clip,width=\textwidth]{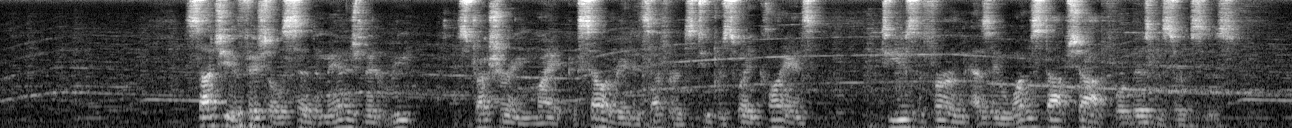}
    \caption{Noisy speech input $\bm{x}$}
    \label{fig:fsegan_x}
  \end{subfigure}%
  ~
  \begin{subfigure}[b]{0.47\linewidth}
 	\centering
    \includegraphics[trim={220 5 910 20},clip,width=\textwidth]{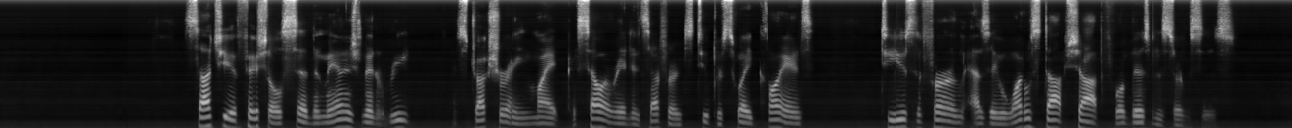}
    \caption{L1-trained output $\mathg(\bm{x}$)}
    \label{fig:fsegan_x}
  \end{subfigure}
  \\
  \begin{subfigure}[b]{0.47\linewidth}
 	\centering
    \includegraphics[trim={220 5 910 20},clip,width=\textwidth]{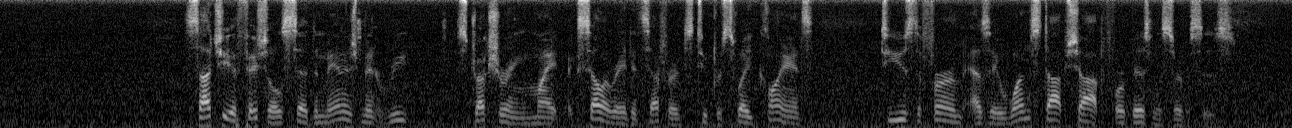}
    \caption{Clean speech target $\bm{y}$}
    \label{fig:fsegan_y}
  \end{subfigure}%
  ~
  \begin{subfigure}[b]{0.47\linewidth}
 	\centering
    \includegraphics[trim={220 5 910 20},clip,width=\textwidth]{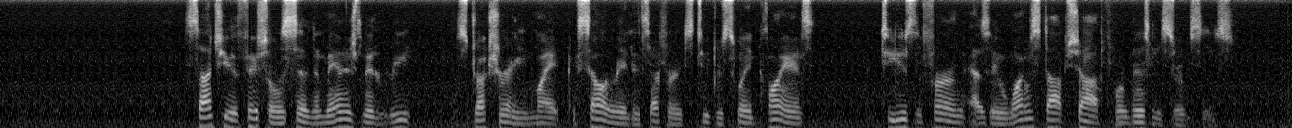}
    \caption{FSEGAN output $\mathg(\bm{x}$)}
    \label{fig:fsegan_Gx}
  \end{subfigure}

  \caption{Noisy utterance enhanced by FSEGAN.}
  \vspace{-0.2in}
  \label{fig:fsegan_vis}
\end{figure}

We show a visualization of FSEGAN enhancement in Figure~\ref{fig:fsegan_vis}.
The procedure appears to reduce both the presence of additive noise and reverberant smearing.
Despite this,
the procedure degrades performance of ASR-MTR.
We hypothesize that the enhancement process may be introducing hitherto-unseen distortions that compromise performance.


\subsection{Retraining Experiments}

Hoping to improve performance beyond that of MTR training alone,
we retrain ASR-MTR using FSEGAN-enhanced features.
To examine the effectiveness of the adversarial component of FSEGAN,
we also experiment with training the same enhancement model using only the L1 portion of the hybrid loss function ($\mathcal{L}_{L1}(\mathg)$ from Equation~\ref{eq:pix2pix}).

Considering that the model may benefit from knowledge of both the enhanced \emph{and} noisy features,
we also train a model to ingest these two representations stacked along the channel axis. 
We initialize this new hybrid model from the existing ASR-MTR checkpoint, 
setting the additional parameters to zero to ensure identical performance at the start of training.
To ensure that the hybrid model is not \emph{strictly} benefiting from increased parametrization,
we train an LAS model from scratch with stereo MTR input.
Results for these experiments appear in Table~\ref{tab:results_retrain}.

\begin{table}[!t]
\small
\centering

\begin{tabular}{lr}

    \toprule
    Model				                    & WER (\%)      \\
    \midrule
    MTR Baseline~* 	                       	& $20.3$    \\
   	~~+ Stereo		 	                	& $19.0$    \\
    \tabvspace
    MTR + FSEGAN Enhancer~*		        	& $25.4$    \\
   	~~+ Retraining		                	& $21.0$    \\
   	~~+ Hybrid Retraining		            & $17.6$    \\
    \tabvspace
    MTR + L1-trained Enhancer~*		        & $21.4$    \\
   	~~+ Retraining		                	& $18.0$    \\
   	~~+ Hybrid Retraining		            & $17.1$    \\
    \bottomrule

\end{tabular}
\caption{Results of ASR-MTR retraining. Rows marked with * are the same model under different enhancement conditions.}
\label{tab:results_retrain}
\vspace{-0.1in}
\end{table}


Retraining ASR-MTR with FSEGAN-enhanced features improves performance by \wer{17} relative to naively feeding them,
but still falls short of MTR training.
Hybrid retraining with both the original noisy and enhanced features improves performance further,
exceeding the performance of stereo MTR training alone by
\wer{7} relative. 
Our results indicate that training the same enhancer with the L1 objective achieves better ASR performance than an adversarial approach,
suggesting limited usefulness of GANs in this context.

\section{Conclusions}

We have introduced FSEGAN, 
a GAN-based method for performing speech enhancement in the frequency domain, 
and demonstrated improvements in ASR performance over a prior time-domain approach.
We provide evidence that,
with retraining,
FSEGAN can improve the performance of existing MTR-trained ASR systems.
Our experiments indicate that, 
for ASR,  
simpler regression approaches may be preferable to GAN-based enhancement. 
FSEGAN appears to produce plausible spectra and may be more useful for telephonic applications if paired with an invertible feature representation.

\section{Acknowledgements}

The authors would like to thank 
Arun Narayanan, 
Chanwoo Kim, 
Kevin Wilson, 
and 
Rif A. Saurous 
for helpful comments and suggestions on this work.

\small
\bibliographystyle{IEEEbib}
\bibliography{refs}

\end{document}